\documentstyle[epsfig,12pt]{article}

\newcommand{\bq}{{\bf q}}
\newcommand{\br}{{\bf r}}

\begin{document}
\title{Many-particle nucleon-nucleon forces from nuclear
single-particle states }
\author{B.L. Birbrair\footnote{Birbrair@thd.pnpi.spb.ru} and
V.I. Ryazanov\\
Petersburg Nuclear Physics Institute\\
Gatchina, St.Petersburg 188350, Russia } \date{} \maketitle

 \begin{abstract}
As follows from the energies of single-particle states in
$^{40}$Ca, $^{90}$Zr and $^{208}$Pb nuclei the contribution of
many-particle $NN$ forces to the nuclear single-particle
potential is at least the sum of repulsive and attractive parts
resulting from three-particle and four-particle forces
respectively.  In addition the specified nucleon density
distributions in the above nuclei are determined from both the 1
GeV proton-nucleus elastic scattering and the single-particle
energies.
\end{abstract}

\section{Introduction}

The nucleon-nucleon interaction proceeds via the meson exchange,
and therefore the existence of many-particle $NN$ forces is a
natural consequence of nonlinearity of the strong interaction
theory. Indeed, only two-particle forces, Fig.1a, are possible
in the lowest order of linear theory containing only quadratic
terms of the meson field $\varphi$ in the meson Lagrangian
density. The situation is different in nonlinear theories. In
the theory with $\varphi^3$ terms the meson may turn into two
ones thus giving rise to three-particle forces, Fig.1b, in
addition to the pairing ones. In the same way the lowest-order
four-particle forces, Fig.1c, appear in the theory with
$\varphi^4$ terms. No higher-power terms exist in renormalizable
theories, but the branching of meson is possible in higher
orders thus giving rise to higher $NN$ forces, Fig.1d. For this
reason the empirical information on many-particle forces is of
fundamental importance for the physics of strong interactions.

At present the only such information is provided by the
calculations for the few-nucleon systems. According to these
calculations something should be added to the two-particle $NN$
forces to get the agreement  between the calculated and observed
quantities. In practice only three-particle forces are added
(see Ref.\cite{1} and the references therein) since the problem
is difficult even in this case. Indeed, the forces are
characterized by a number of adjustable parameters describing
the strength, range, the spin-isospin structure etc., and the
observed quantities are expressed through these parameters in
all orders of the perturbation theory. In such conditions it is
hardly possible to distinguish, for instance, between the
genuine three-particle interaction and the combination of the
latter with the four-particle one.

As shown in the present paper such possibility may be provided
by the experimental data on nuclear single-particle states. It
is based on the facts that (a)~the contribution of the
two-particle $NN$ forces to the nuclear single-particle
potential is fixed by experiment and (b)~no correlation effects
contribute to the single-particle energies \cite{2}. For these
reasons the many-particle contribution may be detected by
comparing the observed energies of single-particle states with
the calculations including the two-particle $NN$ forces only.

\section{Nuclear single-particle states}

We are using the untraditional approach by M.Baranger \cite{2}.
It is based on the spectral representation for the
single-particle propagator \cite{3,4}
\begin{eqnarray}
&& S_A(x,x';\tau)\ =\ -i\langle A_0|T\psi(x,\tau)\psi^+
(x',0)|A_0\rangle  \nonumber \\
= && i\theta(-\tau)\sum_j\psi_j(x)\psi^+_j(x')e^{-iE_j\tau}
-i\theta(\tau)\sum_k\psi_k(x)\psi^+_k(x')e^{-iE_k\tau}\ ,
\end{eqnarray}
where $|A_0\rangle$ is the ground-state wave function of nucleus
$A$ which is chosen to be even-even the ground state thus being
nondegenerate,
\begin{eqnarray}
&& \psi_j(x)\ =\ \langle(A-1)_j|\psi(x)|A_0\rangle\ , \quad
E_j=\ {\cal E}_0(A)-{\cal E}_j(A-1) \nonumber\\
&& \psi_k(x)\ =\ \langle A_0|\psi(x)|(A+1)_k\rangle\ , \quad
E_k=\ {\cal E}_k(A+1)-{\cal E}_0(A)
\end{eqnarray}
the sums over
$j$ and $k$ thus running over complete sets of states of $A-1$
and $A+1$ nuclei. The short-time behaviour of a particle or hole
which is suddenly created in the ground state $|A_0\rangle$ is
described by the following relations
 \begin{eqnarray}
&&\sum_j\psi_j(x)\psi^+_j(x')+\sum_k\psi_k(x)\psi^+_k(x') =
i\bigg(S_A(x,x';+0)-S_A(x,x';-0)\bigg)\nonumber\\
=&& \delta(x-x')\\
&&\sum_jE_j\psi_j(x)\psi^+_j(x')+\sum_kE_k\psi_k(x)\psi^+_k(x')
=-\bigg(\dot S_A(x,x';+0)-\dot S_A(x,x';-0)\bigg) \nonumber\\
=&& H_{sp}(x,x')\\
&&\sum_jE^2_j\psi_j(x)\psi^+_j(x')+\sum_kE^2_k\psi_k(x)\psi^+_k(x'
) = -i\bigg(\ddot S_A(x,x';+0)-\ddot S_A(x,x';-0)\bigg)\nonumber\\
=&& H^2_{sp}(x,x')+\Pi(x,x')\ ,
\end{eqnarray}
where $\dot S=\partial S/\partial\tau$, $\ddot S=\partial^2S/
\partial\tau^2$. The single-particle states are treated as
eigenstates of the single-particle Hamiltonian $H_{sp}(x,x')$,
Eq.(4),
\begin{equation}
\varepsilon_\lambda\psi_\lambda(x)\ =\ \int H_{sp}(x,x')
\psi_\lambda(x')dx'
\end{equation}
thus being the doorway states for the single-nucleon transfer
reactions\cite{2}. As a result of the correlation effects these
states are distributed over actual states of both the $A-1$ and
$A+1$ nuclei. However the properties of the distribution process
permit the determination of the single-particle energies from
the experimental data. Introducing the spectroscopic factors
\begin{equation}
s^{(\lambda)}_{j,k}\ =\ \left|\int\psi^+_\lambda(x)\psi_{j,k}(x)
dx\right|^2\ ,
\end{equation}
multiplying Eqs. (3)--(5) by $\psi^+_\lambda(x)\psi_\lambda(x')$
and integrating over $x$ and $x'$  we get
$$ \sum_js^{(\lambda)}_j+\sum_ks_k^{(\lambda)}\ =\ 1 \eqno{(3a)}
$$
$$ \sum_jE_js^{(\lambda)}_j+\sum_kE_ks_k^{(\lambda)}\ =\
\varepsilon_\lambda \eqno{(4a)} $$
$$ \sum_jE^2_js^{(\lambda)}_j+\sum_kE_k^2s_k^{(\lambda)}\ =\
\varepsilon^2_\lambda+\sigma^2_\lambda \eqno{(5a)} $$
\begin{equation}
\sigma^2_\lambda\ =\ \int\psi^+_\lambda(x)\Pi(x,x')\psi_\lambda
(x')dxdx'\ .
\end{equation}
The determination of single-particle energies by using the Eqs.
(3a)--(5a) and (8) will be discussed in Subsect.3.2.

The Ref.\cite{2} derivation of the explicit form of the
single-particle Hamiltonian $H_{sp}$ should be specified by
taking into account the meson-exchange nature of the $NN$
interaction.

First, the contemporary meson-exchange forces such as the OBE
\cite{5}, Paris \cite{6} and Bonn \cite{7} have no hard
repulsive core, and therefore the problem of elimination of
short-range correlations (see the discussion in Ref.\cite{2})
does not really exist. So there are no actual reasons preventing
from a direct use of bare $NN$ forces for nuclear structure
calculations (i.e. the preliminary calculation of the Brueckner
$G$-matrix \cite{8} or some different way of the hard core
elimination is not necessary).

Second, the commutator technique of Ref.\cite{2} does not apply
in this case because of the retardation. For this reason the
field-theoretical approach \cite{3,4} should be used instead. In
this approach the single-particle Green function
\begin{equation}
G_A(x,x';\varepsilon)\ =\ \int S_A(x,x';\tau)\
e^{i\varepsilon\tau}\ d\tau
\end{equation}
obeys the Dyson equation
\begin{equation}
\varepsilon G_A(x,x';\varepsilon)\ =\ \delta(x-x')+\hat k_x
G_A(x,x';\varepsilon)+\int M(x,x_1;\varepsilon)G_A(x_1,x';
\varepsilon)\ dx_1\ ,
\end{equation}
where $\hat k_x$ is the kinetic energy and the mass operator
$M(x,x';\varepsilon)$ consists of the energy-independent part
$U(x,x')$, which is just the nuclear single-particle potential,
and the energy-dependent one $\Sigma(x,x';\varepsilon)$ which is
responsible for the correlation effects:
\begin{equation}
M(x,x';\varepsilon)\ =\ U(x,x')+\Sigma(x,x';\varepsilon)\ .
\end{equation}
As shown in Refs.\cite{4,9} the quantity
$\Sigma(x,x';\varepsilon)$ vanishes in the $\varepsilon
\to\infty$ limit
\begin{equation}
\Sigma(x,x';\varepsilon)\ =\ \frac{\Pi(x,x')}\varepsilon+\cdots
\end{equation}
(the dots in the rhs are
the higher-power terms of $\varepsilon ^{-1}$) and therefore
\begin{equation}
U(x,x')\ =\ \lim_{\varepsilon\to\infty}\ M(x,x';\varepsilon)
\end{equation}
the decomposition (11) thus being unambiguous.

Let us show that the single-particle Hamiltonian of Eq.(4) is
\begin{equation}
H_{sp}(x,x')\ =\ \hat k_x\delta(x-x')+U(x,x')\ ,
\end{equation}
whereas the quantity $\Pi(x,x')$ of Eq.(5) is defined by
Eq.(12). First let us note that according to the time-energy
Heisenberg relation the infinite $\varepsilon$ value is
equivalent to the infinitely short time interval, the Eqs.(13)
and (14) thus meaning that the Hamiltonian $H_{sp}$ is indeed
responsible for the short-time behaviour of the particle
(hole).

Now let us use the spectral representation \cite{3,4}
\begin{equation}
G_A(x,x';\varepsilon)\ =\ \sum_j\frac{\psi_j(x)\psi^+_j(x')}{
\varepsilon-E_j-i\delta}+\sum_k\frac{\psi_k(x)\psi^+_k(x')}{
\varepsilon-E_k+i\delta}
\end{equation}
and the identity
\begin{equation}
\frac1{\varepsilon-E}\ =\ \frac1{\varepsilon(1-E/\varepsilon)}\
=\ \frac1\varepsilon+\frac E{\varepsilon^2}+
\frac{E^2}{\varepsilon^3} +\ \cdots
\end{equation}
in the $\varepsilon\to
\infty$ limit. Putting Eq.(16) into Eq.(15) we get the following
asymptotic expansion \begin{equation} G_A(x,x';\varepsilon)\ =\
\frac{I_0(x,x')}\varepsilon +
\frac{I_1(x,x')}{\varepsilon^2}+\frac{I_2(x,x')}{\varepsilon^3}
+\ \cdots
\end{equation}
the quantities $I_0(x,x'), I_1(x,x')$  and $I_2(x,x')$ being
just the lhs of Eqs. (3),(4) and (5) respectively. As seen from
Eqs.(11),(12) and (14) the Dyson equation (10) may be written in
the form
\begin{equation}
\varepsilon G_A(x,x';\varepsilon)\ =\ \delta(x-x')+\int
\left(H_{sp}
(x,x_1)+\frac{\Pi(x,x_1)}\varepsilon+\cdots\right)G_A(x_1,x';
\varepsilon)dx_1\ .
\end{equation}
Putting Eq.(17) into Eq.(18) we again get Eqs.(3)--(5), but now
the quantities $H_{sp}$ and $\Pi$ are determined.

In terms of bare $NN$ forces (for a moment let us take into
account the pairing forces only) the single-particle potential
is defined by the first-order diagrams of Fig.2 provided the
nucleons interact without the retardation \cite{9}. This is
however not the case for the meson-exchange forces including
both the momentum and the energy transfer. As a result of the
latter the exchange diagrams of Fig.2b have the
$\varepsilon^{-1}$ asymptotics thus contributing to the quantity
$\Sigma(x,x';\varepsilon)$ rather than $U(x,x')$ \cite{10}. Let
us illustrate this for the Bonn $B$ potential \cite{7}. It is
the sum of the terms
\begin{equation}
v(q,\omega)\ =\ g^2\left(\frac{\Lambda^2-\mu^2}{\Lambda^2+q^2
-\omega^2}\right)^{2\alpha}\ \frac1{\mu^2-q^2-\omega^2}
\end{equation}
in the four-momentum space, the form of the meson-nucleon
vertices being specified by the Lorentz symmetry of the mesons.
The latter is however disregarded because it is irrelevant for
the energy dependence. With this remark the expression for the
Fig.2b diagram becomes
\begin{equation}
M_e(x,x';\varepsilon)\ =\ \int\frac{d^3\bq}{(2\pi)^3}\
e^{i\bq(\br-\br')}\int\frac{id\omega}{2\pi}\ v(q,\omega)
G_A(x,x';\varepsilon+\omega)\ .
\end{equation}
Considering the case of the monopole formfactor, $\alpha=1$, and
using the identity
\begin{eqnarray}
&& \hspace{-1cm}
\left(\frac{\Lambda^2-\mu^2}{\Lambda^2+q^2-\omega^2}\right)^2
\frac1{\mu^2+q^2-\omega^2}\ =\ -\lim_{\delta\to0}\left\{
\frac1{2\omega_\mu(q)}\left(\frac1{\omega-\omega_\mu(q)+i\delta}
-\frac1{\omega+\omega_\mu(q)-i\delta}\right)\right.\nonumber\\
&-&\left.
\left(1-(\Lambda^2-\mu^2)\frac\partial{\partial\Lambda^2}\right)
\frac1{2\omega_\Lambda(q)}\left(\frac1{\omega-\omega_\Lambda(q)
+i\delta}-\frac1{\omega+\omega_\Lambda(q)-i\delta}\right)\right\}\
 ,\\
&& \omega_\mu(q)=(\mu^2+q^2)^{1/2}\ ,\ \quad \omega_\Lambda(q)
=(\Lambda^2+q^2)^{1/2}  \nonumber
\end{eqnarray}
we get
\begin{eqnarray}
&&\hspace{-1cm}
M_e(x,x';\varepsilon) = g^2\int\frac{d^3\bq}{(2\pi)^3}
e^{i\bq(\br-\br')}\left\{\frac1{2\omega_\mu(q)}\left[\sum_j
\frac{\psi_j(x)\psi^+_j(x')}{\varepsilon-E_j+\omega_\mu(q)}
+\sum_k\frac{\psi_k(x)\psi^+_k(x')}{\varepsilon-E_k
-\omega_\mu(q)}\right]\right. \nonumber\\
&& \hspace{-1cm}-\left.
\left(1-(\Lambda^2-\mu^2)\frac\partial{\partial\Lambda^2
(q)}\right)\frac1{2\omega_\Lambda(q)}\left[\sum_j
\frac{\psi_j(x)\psi^+_j(x')}{\varepsilon-E_j+\omega_\Lambda(q)}
+\sum_k\frac{\psi_k(x)\psi^+_k(x')}{\varepsilon-E_k
-\omega_\Lambda(q)}\right]\right\} .
\end{eqnarray}
The sign is also irrelevant because the Bonn $B$ potential is
the sum of terms with different signs. In the $\varepsilon\to
\infty$ limit we get
\begin{equation}
M_e(x,x';\varepsilon)\ =\
\frac{g^2}{4\pi^2}\frac{\delta(x-x')}
\varepsilon\int\limits^\infty_0 q^2\left[\frac1{\omega_\mu(q)}
-\frac1{\omega_\Lambda(q)}-\frac{\Lambda^2-\mu^2}{2\omega^3_
\Lambda(q)}\right]dq\ .
\end{equation}
The integral is convergent because the integrand has the
$q^{-3}$ asymptotics. In addition to Fig.2b the quantity
$\Sigma(x,x';\varepsilon)$ includes an infinite sum of
higher-order Feynman diagrams describing all kinds of the
correlation effects (Pauli, particle--particle, particle--hole,
ground--state etc.).

Thus the only contribution to the nuclear single-particle
potential is provided by the Hartree diagrams of Fig.2a and the
Hartree-like ones of Fig.3 resulting from the many-particle $NN$
forces, the single-particle energies thus being free of the
correlation effects. The latter is the feature of the
Ref.\cite{2} approach, thus permitting the model-independent
studies of nuclear structure.

\section{Results}
\subsection{Single-particle potential}

The Fig.2a contribution to the single-particle potential is the
convolution of the two-particle $NN$ forces with the nucleon
density distribution in nucleus. The latter is determined by the
combined analysis of electron-nucleus and 1 GeV proton-nucleus
elastic scattering data \cite{11}. The two-particle forces are
determined from the deuteron properties and the elastic $NN$
scattering phase shifts at the energies below the pion
production threshold. In this way the "pairing" contribution to
nuclear single-particle potential is fixed by the experimental
data thus being independent of any nuclear model.

We have chosen the Bonn $B$ potential \cite{7} for the
two-particle forces, the motivation being discussed in Sect.4.
This choice enables us to check the status of nuclear relativity
\cite{12} by calculating the scalar and vector fields in nuclear
interior. Using the parameters of Table 5 from Ref.\cite{7} and
the equilibrium nuclear matter density value $\rho_{eq}=0.17$
fm$^{-3}$ we get \cite{10}
\begin{equation}
V\ =\ +284\mbox{ MeV }, \qquad S\ =\ -367\mbox{ MeV }.
\end{equation}
These values are close to those provided by the Dirac
phenomenology \cite{13}. In such conditions it is reasonable to
treat the single-particle wave functions $\psi_\lambda(x)$ as
Dirac bispinors obeying the Dirac equation. The single-particle
Hamiltonian resulting from the Bonn $B$ potential is \cite{10}
\begin{equation}
H_{sp}=\ -i\gamma^0\mbox{\boldmath$\gamma\nabla$}+i\Phi(r)\
\mbox{\boldmath$\gamma$}
\frac{\br}r+(\gamma^0-1)m+V(r)+\gamma^0S(r)\ ,
\end{equation}
$m$ is the free nucleon mass, $\gamma^0$ and {\boldmath$\gamma$}
are Dirac matrices. The scalar field $S(r)$ consists of the
isoscalar and isovector parts resulting from the exchange by
$\sigma$ and $\delta$ mesons. In addition to the isoscalar and
isovector parts resulting from the exchange by $\omega$ and
$\rho$ mesons the vector field $V(r)$ includes the Coulomb
potential. The small isovector quantity $\Phi(r)$ results from
the tensor part of the $\rho$ meson-nucleon coupling. The
details of the calculations are described in Ref.\cite{10}. They
also may be found in the extensive literature on the Walecka
model \cite{12}, see for instance Ref.\cite{14}.

The Dirac equation $H_{sp}\psi_\lambda=\varepsilon_\lambda
\psi_\lambda$ is equivalent to Schr\"odinger-like one for a
particle with the effective mass
\begin{equation}
M(r)\ =\ m+\frac12\bigg(S(r)-V(r)\bigg)
\end{equation}
in the central
\begin{equation}
U(r)\ =\ V(r)+S(r)
\end{equation}
and spin-orbit potentials \cite{15}
\begin{equation}
U_{\ell s}\ =\ \frac1r\ \frac d{dc}\left(\frac1{2M(r)}\right)\
\mbox{\boldmath$\ell\sigma$}\ .
\end{equation}
The terms arising from the quantity $\Phi(r)$ are omitted here,
but actually they are taken into account in the calculations.
The quantities in Eqs.(24)--(28) should be supplied by the
subscript "pair" since they describe the contribution from the
two-particle forces.

The contribution from the many-particle forces is looked for as
the following expansion in terms of the nucleon density
distribution $\rho(r)$
\begin{equation}
U_m(r)\ =\ \hbar c\left[a_3\rho^2(r)+a_4\rho^3(r)+\
\cdots\right]
\end{equation}
the $\rho^2$ term results from three-particle forces, Fig.3a,
the $\rho^3$ one is of four-particle origin, Fig.3b, etc. (the
linear density terms enter the two-particle contribution,
Eq.(27)). The potential $U_m(r)$ is assumed to be equally
distributed between the scalar and vector fields,
\begin{equation}
S_m(r)\ =\ V_m(r)\ =\ \frac12\ U_m(r)\ .
\end{equation}
The only motivation for this assumption is the aim to have as
little free parameters as possible.

\subsection{Single-particle energies}

As mentioned in Sect.2 the single-particle energies may be
determined from experiment by using the sum rules (3a)--(5a).
The most suitable situation is that for the cases when the
absolute value of the single-particle energy $\varepsilon_
\lambda$ exceeds the width $\sigma_\lambda$ of the distribution
region. In these cases all states, over which the
single-particle one is distributed, belong to the same nucleus,
and therefore the sum rules (3a)--(5a) are saturated by only one
term in the lhs, the first for the hole states and the second
for the particle ones. Such situation occurs for the peaks in
the cross sections of quasielastic knockout reactions $(p,2p)$
and $(p,pn)$ \cite{16} leading to the hole states with
\begin{equation}
|\varepsilon_\lambda|\ >\ \sigma_{\,ax}\ \cong\ 20 \mbox{ MeV }.
\end{equation}
Indeed, according to Eq.(8) the width $\sigma_\lambda$ depends
on the single-particle wave function $\psi_\lambda$ rather than
the energy $\varepsilon_\lambda$, thus being roughly the same
for all single-particle states. So it is reasonable to identify
$\sigma$ with the largest observed value $\sigma_{\max}\cong20$
MeV. The latter is the width of the peaks corresponding to the
$1s_{1/2}$ hole states. For these reasons the average energies
of the above peaks obeying the Eq.(31) condition may be
identified with the single-particle energies within the
experimental accuracy of $2\div3$ MeV. In this way we
demonstrated that the Ref.\cite{2} approach permits the
model-independent determination of the single-particle energies.
It is worth mentioning that the experimental data for the $sp$
energies are independent of those for the two-particle
contribution to the $sp$ potential.

We used the facts that the cross section of the quasielastic
knockout reaction leading to the fixed nuclear state is
proportional to the spectroscopic factor of this state, and the
absolute values of the $s$-factors are not necessary when all
states, over which the single-particle one is distributed,
belong to the same nucleus (in this case the relative values are
sufficient). This is however not the case for weakly bound
single-particle states with
$|\varepsilon_\lambda|<\sigma_{\max}$. Such states are
distributed over actual ones of both $A-1$ and $A+1$ nuclei, and
therefore the $s$-factors from both the pickup and stripping
reactions are necessary. But the $s$-factors are determined with
a rather low accuracy because of both experimental and
theoretical ambiguities, and therefore it is unclear how to pin
together the $s$-factors from the two reactions. For these
reasons the energies of weakly bound single-particle states are
yet unknown. One should also bear in mind that the low-lying
states of $A\pm1$ nuclei forming the Fermi-surface of the
closed-shell nucleus $A$ are Landau--Migdal quasiparticles
\cite{4} rather than the single-particle states of nucleon.
Indeed, the correlation term $\Sigma(x,x';\varepsilon)$ is
included in the quasiparticle energies in contrast to the
single-particle ones, see Ref.\cite{4} for details.

\subsection{Many-particle forces and specified density
distributions}

The observed energies of neutron and proton single-particle
states in $^{90}$Zr are plotted in Fig.4 together with the
results of the calculations. As seen from the figure  the
lowest "pair" states $1s_{1/2}$ are significantly underbound
whereas the higher states, especially the $2s_{1/2}$ ones, are
overbound. The same compression of "pair" single-particle states
occurs in $^{208}$Pb, Fig.5, where this effect is even more
pronounced, and also in $^{40}$Ca, Fig.6. This means that the
potential well, including two-particle forces only, is somewhat
too wide but insufficiently deep. The isoscalar potentials in
$^{90}$Zr, "pair", actual and the many-particle contribution, are
plotted in Fig.7. As seen from the figure, the many-particle
contribution (i.e. the difference between the actual and "pair"
wells) consists of the repulsive and attractive parts, the
radius of the latter being less than that of the former. So the
expansion (29) has to contain at least two terms of different
sign obeying the above condition. The most simple possibility is
provided by the sum of the three-particle repulsion, $a_3>0$,
and the four-particle attraction, $a_4<0$. Of course we cannot
guarantee the absence of contributions from higher many-particle
forces.  It only should be mentioned that the above possibility
corresponds to the least number of free parameters.

Taking into account the possible contribution of many-particle
forces to the isovector nuclear potential the quantity $U_m(r)$
is chosen as
\begin{eqnarray} U_m(r) &=& \hbar
  c\bigg\{a_3\rho^2(r)+a_4\rho^3(r)-\tau_3
\left[a^-_3\rho(r)+a^-_4\rho^2(r)\right]\rho^-(r)\bigg\}\ ,
  \nonumber\\
\rho(r) &=& \rho_n(r)+\rho_p(r)\ , \quad \rho^-(r)\ =\
  \rho_n(r)-\rho_p(r)\ .
  \end{eqnarray}
  $\rho_n(r)$ and $\rho_p(r)$ are neutron and proton  density
  distributions in nucleus. As seen from Fig.4 the description
of single-particle energies is improved by including  the
many-particle contribution.

The results, which are labelled as "tot" in Fig.4, are obtained
  using the Woods--Saxon-like density distributions of
Ref.\cite{11} which are folded with the nucleon electromagnetic
form factor, see Ref.\cite{11} for details. However the nuclear
  potential is expressed through the point densities since the
  finite size of nucleon is taken into account in the $NN$
  forces. For this reason we used the point densities
$\rho_0W_A(r)$ which are obtained from those of Ref.\cite{11}
  by usual deconvolution procedure.

It should be mentioned that
  the electron and proton elastic scattering data underlying
  these densities are sensitive to nucleon density distributions
  in the surface region of nucleus, whereas the single-particle
  energies are sensitive to those in nuclear interior. Therefore
  the observed single-particle energies may be used to specify
  the nucleon density distributions. We tried many different
  forms of $\rho(r)$. The most appropriate one is found to be
  \begin{equation}
  \rho(r)\ =\ \rho_0\bigg(W_A(r)+\alpha W_A(0)\varphi_4(\beta r)
  \bigg)\ ,
  \end{equation}
where $\varphi_4(x)$ is the fourth order Hermite function. The
best fit parameters $\alpha$ and $\beta$ of neutron and proton
density distributions are shown in Table 1. The best fit
strength parameters are found to be the same for all nuclei.
They are
\begin{eqnarray}
&& a_3=13.6608\mbox{ fm}^5\ , \quad a_4=-80.2568\mbox{ fm}^8 \\
&& a^-_3=28.7531\mbox{ fm}^5\ , \quad a^-_4=-144.3534\mbox{
  fm}^8\ . \nonumber
  \end{eqnarray}
The calculations with the specified densities are labelled as
  "tot1" in Figs. 4--6.

\begin{table}\caption{
   Neutron and proton density parameters}

\begin{center}
\begin{tabular}{||r|c|c|c|c||} \hline\hline
  & $\alpha_n$ & $\beta_n$ & $\alpha_p$ & $\beta_p$\\
  \hline\hline
  $^{90}$Zr & $-0.0836$ & 0.4823 & $-0.0287$ & 0.4416\\
  $^{40}$Ca & $-0.0513$ & 0.5469 & $-0.0215$ & 0.4557\\
  $^{208}$Pb & $-0.2146$ & 0.4670 & $-0.0092$ & 0.2200\\
    & $-0.3083$ & 0.4646 & $-0.1308$ & 0.2779\\
  \hline\hline
  \end{tabular}\end{center} \end{table}

Slightly better agreement for $^{208}$Pb is provided by the
following strength parameters
\begin{eqnarray}
&& a_3=15.1120 \mbox{ fm}^5, \quad a_4=-90.9870\mbox{ fm}^8\ ,
\nonumber\\
&& a_3^-=21.0000\mbox{ fm}^5, \quad a^-_4=-99.0890\mbox{ fm}^8
\end{eqnarray}
with the following form of the nucleon density distributions
\begin{equation}
\rho(r)\ =\ \rho_0W_A(r)\left(1+\alpha\varphi_4(\beta
r^{})\right)\ .
\end{equation}
The corresponding parameters $\alpha$ and $\beta$ are those of
the fourth row in Table 1, the results are labelled as "tot2" in
Fig.5.

The largest discrepancy between the observed single-particle
energies and the "tot1" results for $^{90}$Zr and $^{40}$Ca as
well as the "tot2" ones for $^{208}$Pb is less than 3 MeV, the
average discrepancy is 1.65 MeV.

The specified nucleon density distributions are plotted in
Figs.8 and 9.  As seen from the figures the neutron densities
have a pronounced dip in the center of nucleus which increases
with increasing mass number.

We also calculated the 1 GeV proton elastic scattering cross
sections within the Glauber--Sitenko theory \cite{17} using both
the specified densities and those of Ref.\cite{11}. The results,
see Fig.10, clearly show that the agreement with experiment
is equally good for both the Ref.\cite{11} and the specified
densities.

The possible reasons for the difference between the Eq.(34) and
the Eq.(35) strength parameters are as follows. (i)~Expression
(32) corresponds to the zero-range forces \begin{eqnarray} &&
f_3(r_{12},r_{13})\ =\ (a_3+\mbox{\boldmath$\tau$}_1
\mbox{\boldmath$\tau$}_2 a^-_3)\delta(r_{12})\delta(r_{13})\\
&& f_4(r_{12},r_{13},r_{14})\ =\ (a_4+\mbox{\boldmath$\tau$}_1
\mbox{\boldmath$\tau$}_2a^-_4)\delta(r_{12})\delta(r_{13})
\delta(r_{14})\ ,
\end{eqnarray}
whereas the actual forces may be of finite range. (ii)~The above
forces should have been folded with the two-particle density and
the three-particle one rather than the products of
single-particle densities. (iii)~As mentioned above, the
contributions from higher many-particle forces may be present.

The difference is however small for the charge-symmetric nuclear
matter with the equilibrium density $\rho_{eq}=0.17$~fm$^{-3}$.
Indeed, as follows from Eqs. (32),(34) and (35) the
contributions of three(four)-particle forces to the isoscalar
single-particle potentials are
\begin{eqnarray}
&& U_3\ =\ \hbar ca_3\rho^2_{eq}\ =\ 78~(86)\mbox{ MeV },\\
&& U_4\ =\ \hbar ca_4\rho^3_{eq}\ =\ -78~(-88)\mbox{ MeV }.
\end{eqnarray}
The first values in the rhs correspond to the Eq.(34) parameters
whereas those in parentheses refer to the Eq.(35) set. They are
not small compared to the "pair" value $U_{\rm pair}=-83$~MeV,
see Eqs. (24) and (27). But they nearly compensate each other
thus giving rise to the conclusion that the isoscalar part of
the nuclear single-particle potential is mainly of the
two-particle origin. This conclusion is supported by the fact
that according to Eq.(30)
\begin{equation}
S_3=V_3=39\,(43)\mbox{ MeV },
\quad S_4=V_4=-39\,(-44)\mbox{ MeV }, \end{equation} thus being
considerably less than the pair values $S_{\rm pair}=-367$~MeV,
$V_{\rm pair}=284$~MeV, see Eq.(24).

The situation is different for the isovector part of the
potential. With  $\rho^-=\frac{N-Z}A\rho_{eq}\ $ Eq.(32) gives
\begin{equation}
U^-_m\ =\ \hbar c\left(a^-_3\rho_{eq}+a^-_4\rho^2_{eq}\right)
\rho^-\ =\ 24\ \frac{N-Z}A\mbox{  MeV }
\end{equation}
for both Eq.(34) and Eq.(35) parameters whereas the "pair" value
provided by the Bonn $B$ potential is $U^-_{\rm
pair}=7\frac{N-Z}A$MeV, the isovector nuclear potential thus
being mainly of many-particle origin. The reason is due to the
fact that the two-particle contribution arises from the exchange
by isovector mesons $\rho$ and $\delta$ which are weakly coupled
to nucleon \cite{7}.

\section{Summary}
In this way we demonstrated that the many-particle $NN$
interaction includes at least the three-particle repulsion and
the four-particle attraction. This result is restricted because
no information is obtained for the many-particle forces which
do not contribute to the nuclear single-particle potential. But
it provides a very instructive example of the situation where
the experimental data on complex nuclei are more appropriate for
the fundamental problem than those on few-nucleon systems.
Indeed, the latter ones are described by solving the
complicated many-particle quantum mechanical problem including
the interaction in all orders of the perturbation theory. On the
contrary the single-particle states are solutions of the
one-particle problem which is much more simple. In addition the
nuclear single-particle potential is expressed through the $NN$
interaction in first order of the perturbation theory, the
result thus being visual (see Fig.7).

Besides the above-mentioned $A$ dependence of the strength
parameters our results may have the following additional
ambiguities.

1. They are essentially based on the choice of the Bonn
potential for the two-particle $NN$ forces. The reason for this
choice is a very high level of confidence: (a)~the physics
underlying the Bonn potential is absolutely clear, (b)~it
contains only one adjustable parameter, (c)~the two-nucleon data
are described with $\chi^2/$datum=1.9 \cite{18} (according to
this reference the Bonn $B$ version we used in the present work
is completely equivalent to the full one).

Of course our results may
be changed with the progress of the knowledge about the $NN$
interaction. We insist however that the acceptable new-fashioned
potential must be of higher level of confidence than the Bonn
one. This means that (a)~the underlying physics  should be as
clear as that for the Bonn potential, (b)~the number of free
parameters may be also only one but the $\chi^2$/datum value
should be considerably less or (c)~the $\chi^2$/datum value may
be the same but no free parameters should be present.

2. In the Ref.\cite{16} experiments the energy of the knocked-out
nucleon is only about 100 MeV. This may be insufficient to
neglect the final-state inelastic interactions leading to
additional excitation of the final nucleus. As a result of such
excitations the average energies of the peaks may be shifted
from the single-particle energy values because the reaction
mechanism is not a pure quasielastic knockout in this case. To
get rid of this ambiguity the additional quasielastic knockout
$(p,p'N)$ or $(e,e'N)$ experiments are desired in which the
energy of the knocked-out nucleon would be about 1 GeV. We hope
that our work will stimulate such experiments.

The authors are indebted to Dr. M.B. Zhalov for valuable
discussions.


\newpage

\subsection*{Figure Captions}

\begin{description}

\item[Fig.1.] Two-particle (a), three-particle (b),
four-particle (c) and higher (d) nucleon-nucleon interaction
forces. The wavy lines are for mesons, the full ones are for
nucleons.

\item[Fig.2.] First-order Hartree (a) and exchange (b) diagrams.

\item[Fig.3.] Contributions of three-particle (a) and
four-particle (b) forces to nuclear single-particle potential.

\item[Fig.4.] Single-particle energies of neutron and proton
states in $^{90}$Zr. The energy scale is shown from the left of
each figure. The labels are $"\exp"$ for the observed energies,
"pair" for the calculations taking into account the two-particle
forces only, "tot" for those including the many-particle forces
and using the density distributions from Ref.\cite{11}, and "tot
1" for the case when both the many-particle forces and specified
densities are included.

\item[Fig.5.]
The same for $^{208}$Pb. The label "tot 2" is for the
case when the individual strength parameters for $^{208}$Pb are
 used.

\item[Fig.6.] The same for $^{40}$Ca.

\item[Fig.7.]
Isoscalar potential in $^{90}$Zr. The full, dashed and
  dot-dashed lines are for actual, "pair" and many-particle part
  of the potential respectively.

\item[Fig.8.] Neutron density distributions in $^{40}$Ca,
$^{90}$Zr and $^{208}$Pb.

\item[Fig.9.] The same for protons.

\item[Fig.10.]
Elastic scattering of 1 GeV protons on $^{40}$Ca, $^{90}$Zr and
$^{208}$Pb.  The calculations with the specified densities are
shown by full line, those with the Woods--Saxon-like ones from
Ref.\cite{11} are plotted by dashed line, the dots are for the
experimental data.

\end{description}


\begin{figure}
\centerline{\vspace{1.0cm}\hspace{0.1cm}\epsfig{file=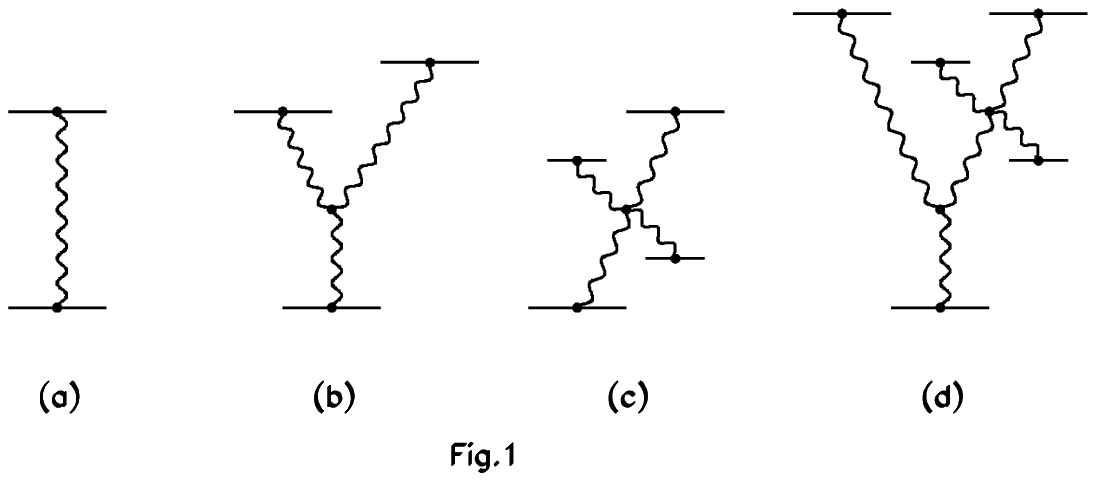,width=12cm}}
\vspace*{1cm}
\centerline{\vspace{1.0cm}\hspace{0.1cm}\epsfig{file=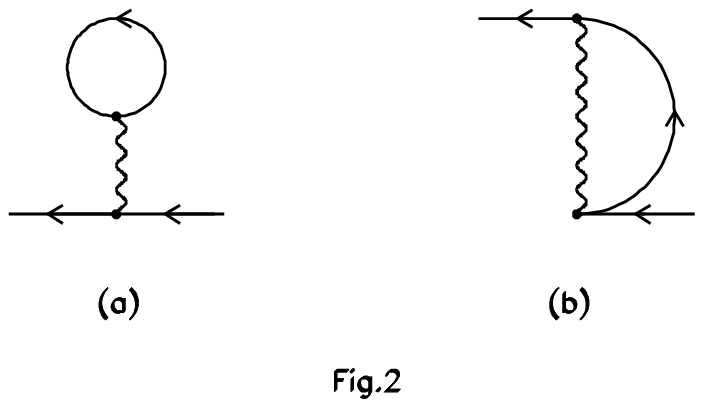,width=10cm}}
\vspace*{1cm}
\centerline{\vspace{1.0cm}\hspace{0.1cm}\epsfig{file=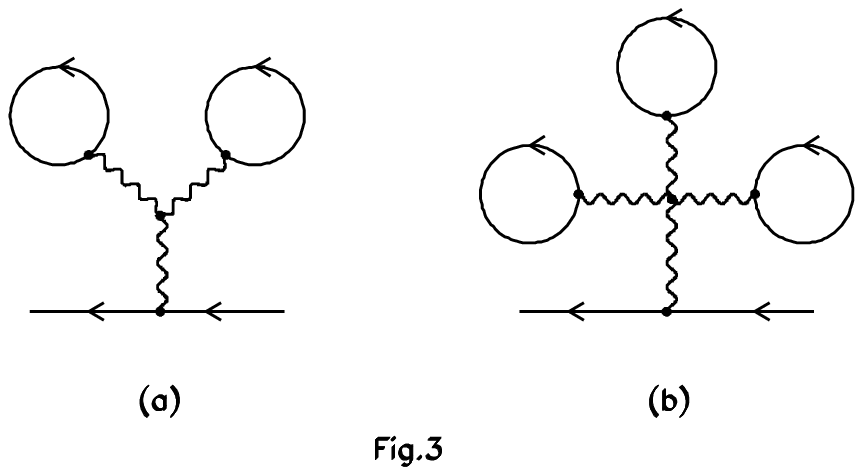,width=10cm}}
\end{figure}

\begin{figure}
\centerline{\vspace{0.2cm}\hspace{0.1cm}\epsfig{file=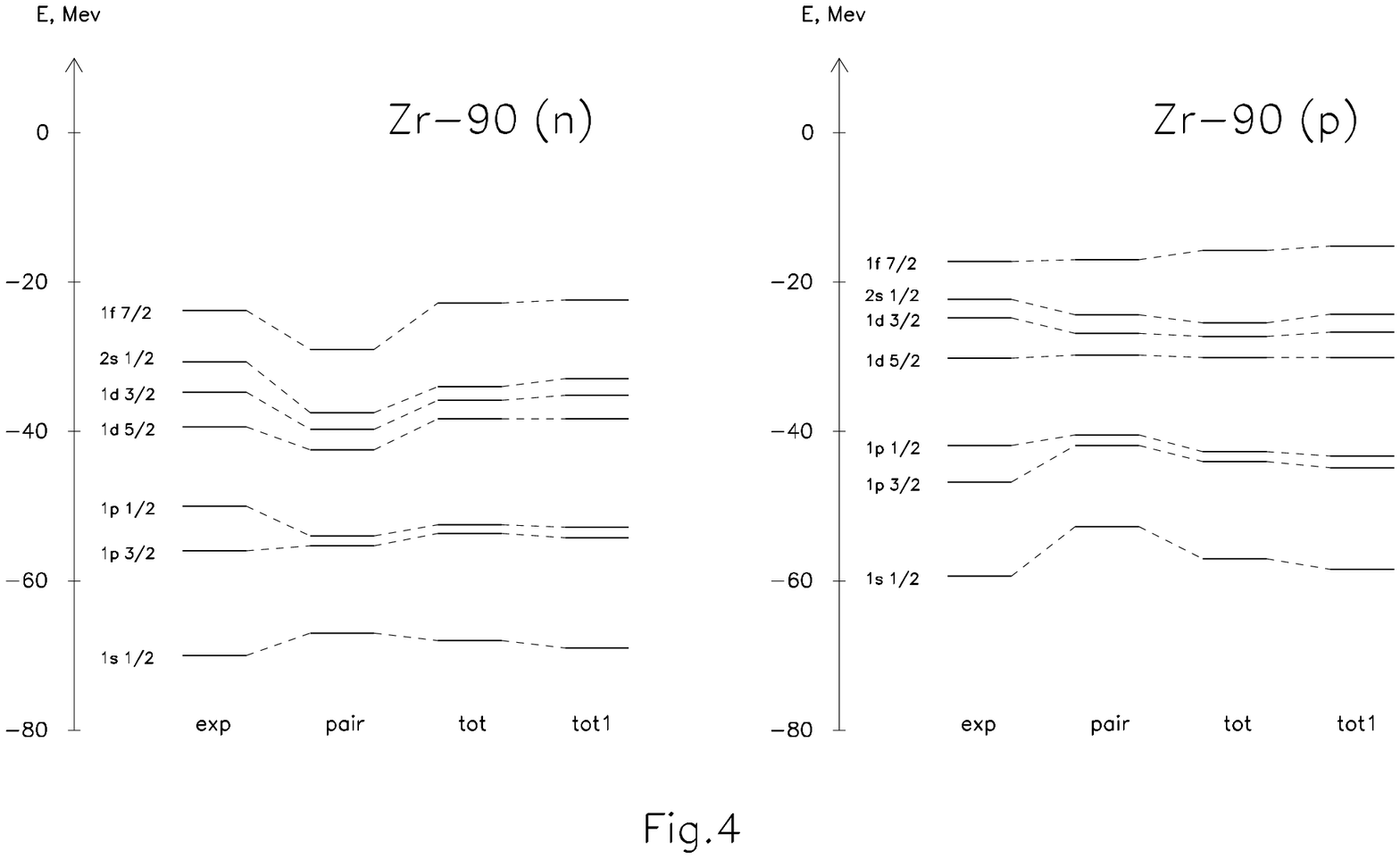,width=10cm}}
\vspace*{0.8cm}
\centerline{\vspace{0.2cm}\hspace{0.1cm}\epsfig{file=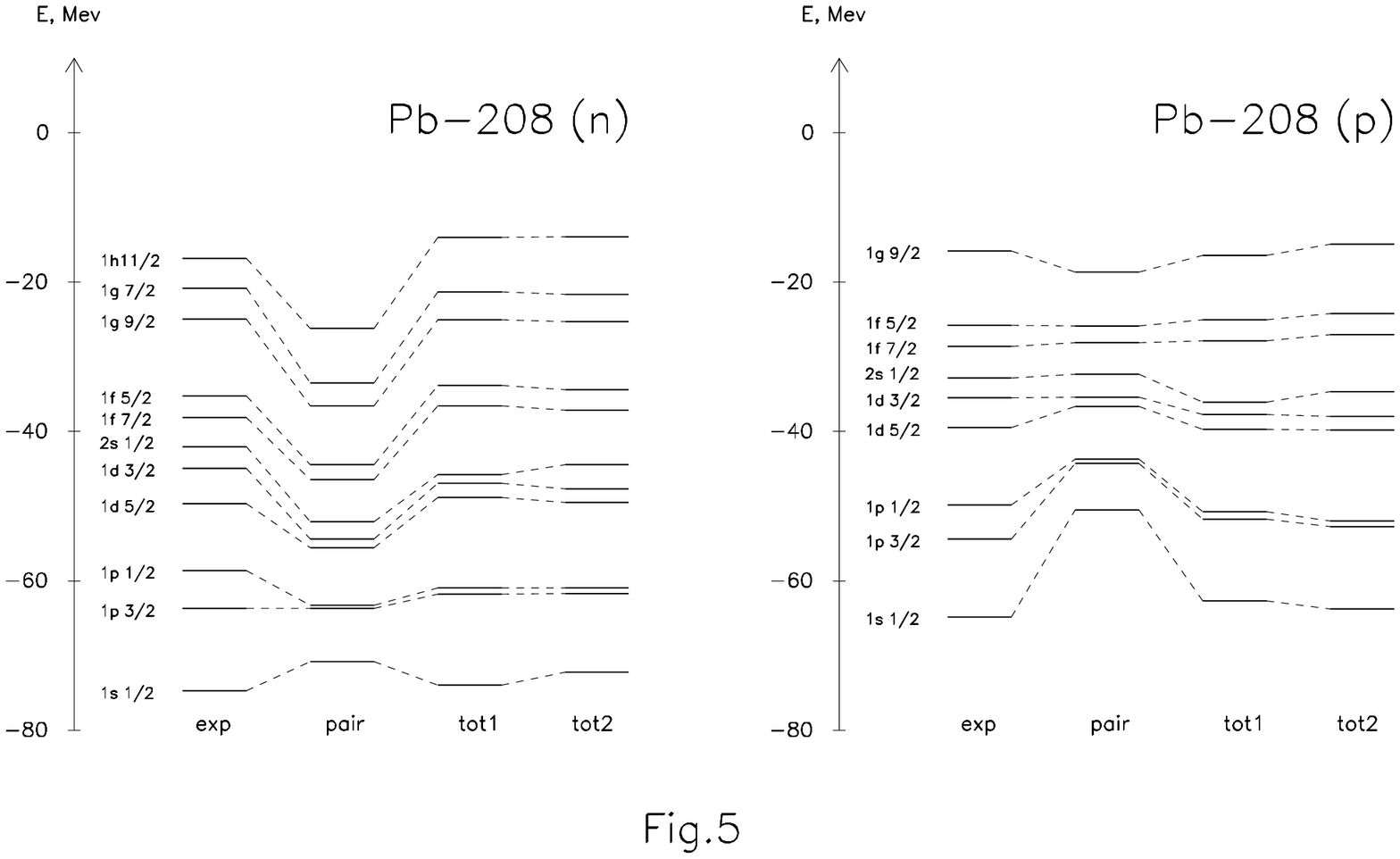,width=10cm}}
\vspace*{0.8cm}
\centerline{\vspace{0.2cm}\hspace{0.1cm}\epsfig{file=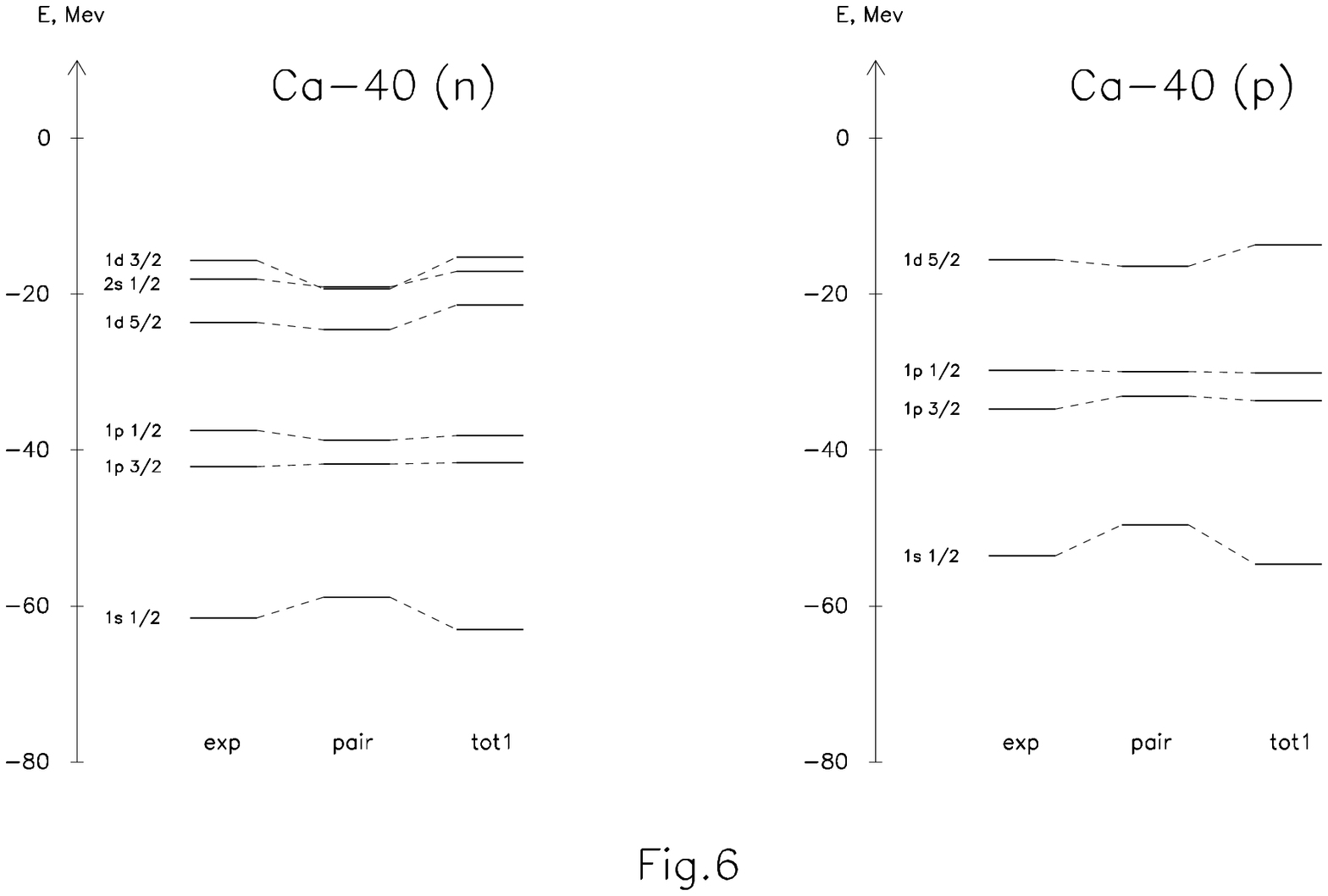,width=10cm}}
\end{figure}

\begin{figure}
\centerline{\vspace{0.2cm}\hspace{0.1cm}\epsfig{file=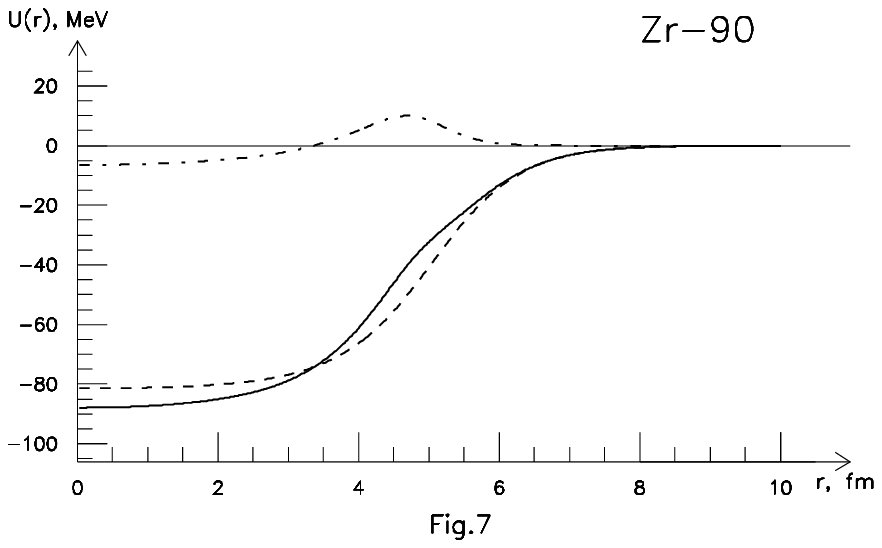,width=9cm}}
\vspace{0.8cm}
\centerline{\vspace{0.2cm}\hspace{0.1cm}\epsfig{file=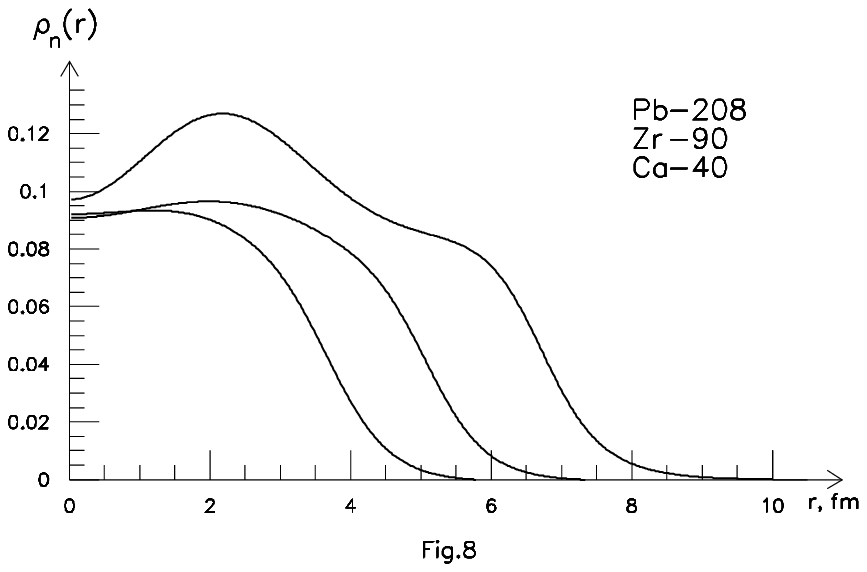,width=9cm}}
\vspace*{0.8cm}
\centerline{\vspace{0.2cm}\hspace{0.1cm}\epsfig{file=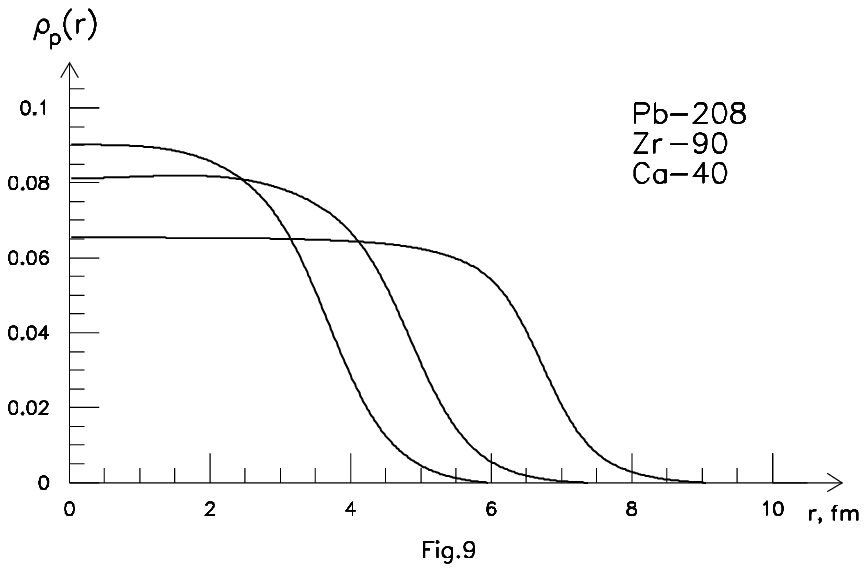,width=9cm}}
\end{figure}

\begin{figure}
\centerline{\vspace{1.0cm}\hspace{0.1cm}\epsfig{file=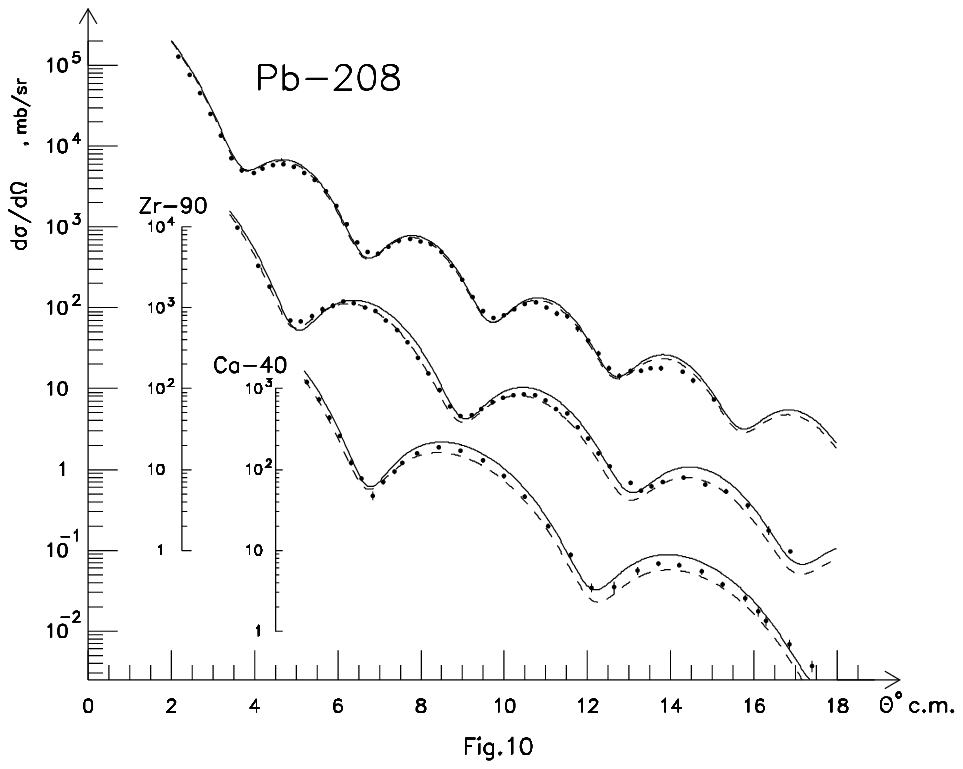,width=12cm}}
\end{figure}


\end{document}